\documentclass[10pt]{article}
\usepackage{decompute-research}

\begin{document}
\thispagestyle{plain}

\begin{HeroPanel}
\includegraphics[width=0.62in]{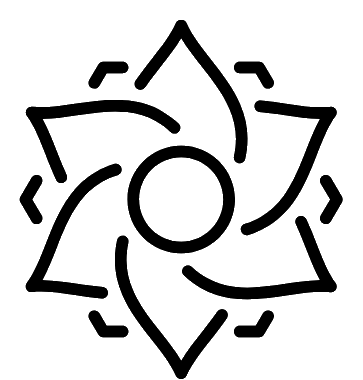}\par\vspace{0.55em}
{\sffamily\bfseries\fontsize{27}{31}\selectfont Echelon: Auditable Aggregate-Only\\[-0.08em] Language-Model Adaptation Across\\[-0.08em] Privacy Boundaries\par}
\vspace{1.05em}
{\sffamily\bfseries Hina Dixit, Punit Kumar, Irene Tenison, Nevasini Sasikumar\par}
\vspace{0.25em}
{\sffamily\textcolor{dcMuted}{Decompute Inc., Cupertino}\par}
\vspace{0.25em}
{\sffamily\footnotesize\textcolor{dcMuted}{Correspondence: \href{mailto:hina@decompute.run}{\texttt{hina@decompute.run}}}\par}
\vspace{0.9em}
\textcolor{dcRule}{\rule{\linewidth}{0.6pt}}
\vspace{0.75em}

{\sffamily\Large\bfseries Abstract}\par\vspace{0.35em}
{\small
Cross-organization language-model adaptation increasingly faces hard governance constraints: in many deployments, device-level model state -- parameters, activations, optimizer state, and per-device updates -- cannot be exported outside an administrative boundary. Existing distributed and federated stacks typically assume cross-site model exchange and then retrofit privacy mechanisms, which complicates compliance and makes auditing brittle.
We present \echelon{}, a boundary-first training architecture that enforces device-level model-state non-export as a systems invariant. Devices train locally inside each boundary; the only cross-boundary payloads are securely aggregated boundary-level deltas plus $O(1)$ coordination metadata, exposed through a concrete audit surface.

Restricting exchange to aggregates changes the optimization problem: the system must remain stable under WAN delay, heterogeneous participation, churn, and non-IID data even though the global plane never sees per-device updates. \echelon{} combines buffered semi-asynchronous secure aggregation, staleness-aware weighting, participation windows, proximal local objectives, and a drift-aware outer synchronization controller.
\\
In 1B-parameter LoRA adaptation across $M=2$ boundaries, a budget-matched contest over three seeds (24.88M tokens) reaches validation loss $3.887\pm 0.010$ and is best or tied-best among tuned low-communication baselines under fixed-token, fixed-bytes, fixed-wall-clock, and fixed-sync-count budgets. In OpenWebText stress tests, \echelon{} sustains $2{,}139$--$2{,}176$ tokens/s across evaluated WAN and non-IID treatments, \echelon{}-DA improves time-to-target under WAN latency relative to a privacy-parity DiLoCo+SA baseline, and quality degrades by at most 2.2\% under 200\,ms emulated latency or severe non-IID partitioning. We further define $\Beff$, a telemetry-derived effective divergence budget, and present a normalized power-law fit as a candidate empirical law for the quality--communication frontier across the evaluated operating window. An audit trace shows zero per-device payload bytes observed under loopback, emulated WAN, and real cross-region deployments.
}
\end{HeroPanel}

\begin{ScopeBox}
{\sffamily\bfseries Core claim and scope}\par
\echelon{} enforces \emph{device-level model-state non-export} as an auditable systems invariant while aiming to preserve near-centralized quality under WAN latency and heterogeneity in the adaptation regimes studied here.
\end{ScopeBox}

\clearpage

\section{Introduction}
Cross-boundary model development is increasingly constrained not just by communication cost, but by governance. In many deployments, organizations may collaborate on a shared model objective while still prohibiting the export of device-level model artifacts outside an administrative, institutional, or jurisdictional boundary. This is stricter than the communication assumptions of conventional distributed training and more operational than privacy phrased only as a statistical guarantee. In these settings, the systems question is not simply how to optimize across sites. It is how to do so while making the allowed information flow explicit, enforceable, and auditable.

Most existing training stacks are not organized around that constraint. Conventional large-scale training systems assume a tightly coupled cluster and allow model state to move freely across nodes. Federated and decentralized methods handle heterogeneity and weak connectivity better, but typically still treat cross-site model exchange as the default substrate, with secure aggregation or other privacy mechanisms layered on top. That design is often reasonable for communication efficiency, but it makes compliance difficult to reason about. The privacy boundary sits outside the training architecture instead of becoming a first-class systems object.

\echelon{} takes the opposite position. It begins from a hard information-flow rule: no per-device parameters, activations, optimizer states, or per-device updates may leave a boundary. Only boundary-level securely aggregated deltas and a small amount of coordination metadata may cross boundaries. This turns structural privacy from an overlay into a systems invariant. The optimizer is then in service of that invariant: once exchange is limited to aggregates, the system must still remain stable under WAN delay, non-IID data, participant churn, and stragglers.

The design has two layers. At the systems layer, \echelon{} organizes execution into device, boundary, and global planes. Devices optimize locally against a boundary reference. Boundaries buffer updates and aggregate them semi-asynchronously with clipping, staleness-aware weighting, and participation windows that limit starvation. The global plane only sees boundary-level aggregates and mixes them at an outer cadence. At the control layer, \echelon{}-DA monitors boundary drift and adapts the outer synchronization interval to observed divergence, synchronizing more frequently when boundaries pull apart and relaxing coordination when they remain aligned.

This paper targets an admissibility-constrained training regime. Under the information-flow rule considered here, methods that export per-device model updates or device-level model state across administrative boundaries are not deployable without additional privacy mechanisms. We therefore distinguish unconstrained performance baselines from privacy-parity baselines. The central question is not whether aggregate-only training outperforms centralized training in an unconstrained cluster, but whether useful adaptation remains possible when the boundary contract is enforced and whether the resulting information flow can be audited.

Together, these components make structural privacy checkable: the boundary contract is explicit, cross-boundary messages are typed, and deployment logs can be audited for violations. The core question we study is whether \echelon{}'s aggregate-only interface still permits effective adaptation under WAN delay, churn, and heterogeneous data. Across the evaluated 1B-parameter LoRA regimes, the answer is yes: \echelon{} remains competitive with tuned low-communication baselines under matched budgets while exposing a small, auditable surface for compliance.

\subsection{Contributions}
\begin{enumerate}
\item \textbf{Boundary-first information-flow contract and audit surface.} We formalize a three-plane training architecture organized around immutable privacy boundaries and an explicit contract: device-level parameters, activations, optimizer state, and per-device updates never cross a boundary; only securely aggregated boundary deltas and $O(1)$ coordination metadata may cross, with typed messages and on-the-wire byte accounting that supports audit.
\item \textbf{Aggregate-only training stack with drift-aware control.} We combine buffered semi-asynchronous secure aggregation, staleness-aware weighting, participation windows, a proximal local objective, and a drift-aware outer synchronization controller to maintain stability under WAN delay, churn, and non-IID data despite aggregate-only exchange.
\item \textbf{Competitive adaptation under matched budgets, WAN stressors, and throughput stress.} In 1B-parameter LoRA adaptation across two boundaries, a budget-matched contest over three seeds reaches validation loss $3.887\pm 0.010$ and is best or tied-best among tuned low-communication baselines under fixed-token, fixed-bytes, fixed-wall-clock, and fixed-sync-count normalizations. In OpenWebText stress tests, \echelon{} sustains $2{,}139$--$2{,}176$ tokens/s across evaluated WAN and non-IID treatments, quality degrades by at most 2.2\% under 200\,ms emulated latency or severe non-IID partitioning, and \echelon{}-DA improves time-to-target under WAN latency relative to a privacy-parity DiLoCo+SA baseline, reducing 100\,ms WAN time-to-target from 145\,min to 95\,min. An end-to-end audit trace over loopback, emulated WAN, and real cross-region deployments confirms that the global plane observes \texttt{per\_device\_payload} of exactly zero bytes.
\item \textbf{Candidate effective-divergence law.} We define an effective divergence budget $\Beff$ from controller telemetry and show that a normalized power-law relation predicts the quality--communication frontier across the evaluated stressors. The result is intentionally scoped: we present $\Beff$ as a candidate empirical law for this training stack and operating window, not as a universal scaling law. Its value is that a single online-computable scalar captures much of the predictive signal of richer stressor models while remaining interpretable and auditable.
\end{enumerate}

\section{Auditability and Evidence Mapping}
Because the goal is auditable boundary-local training, we adopt an explicit claim--evidence posture. For each major claim, the paper identifies (i) the invariant being asserted, (ii) the evidence used to support it, (iii) measurements that would falsify it, and (iv) the limits of the current evidence. Table~\ref{tab:evidence-map} summarizes this mapping.

\begin{table}[H]
\centering
\begin{tighttable}
\caption{Claim-to-evidence map for audit review. ``Status'' distinguishes executed measurements from design-level or forward-looking claims.}
\label{tab:evidence-map}
\begin{tabularx}{\linewidth}{>{\raggedright\arraybackslash}p{0.22\linewidth} >{\raggedright\arraybackslash}X >{\raggedright\arraybackslash}p{0.20\linewidth} >{\raggedright\arraybackslash}p{0.18\linewidth}}
\toprule
Claim & Evidence in manuscript & Main caveat & Status \\
\midrule
Device-level model-state non-export & Information-flow contract (Table~\ref{tab:contract}); schema audit showing zero per-device payload bytes (Table~\ref{tab:env-audit}) & Does not imply differential privacy or malicious-adversary robustness & Executed audit + design invariant \\
Per-round secure aggregation hides individual updates from the coordinator & Secure-aggregation view in Section~\ref{sec:privacy-scope}; quorum-satisfying aggregate is the observable message & Longitudinal leakage remains possible across repeated rounds & Scoped protocol claim \\
Aggregate-only adaptation can remain useful under stress & Budget-matched C4 contest (Tables~\ref{tab:bm-main} and~\ref{tab:bm-budget}); OpenWebText stress results (Table~\ref{tab:wr-stress}) & Current evidence is 1B LoRA, sequence length 32, mostly $M=2$ & Executed experiments \\
Drift-aware control is observable and auditable & Controller trace maps measured drift to synchronization cadence (Table~\ref{tab:controller}) & Trace demonstrates mechanism, not universal optimality & Executed diagnostics \\
$\Beff$ is a candidate effective-divergence law for the evaluated operating window & Normalized power-law collapse and predictive validation (Figures~\ref{fig:beff-quality}, \ref{fig:beff-comm}; Table~\ref{tab:beff-validation}) & Current evidence is 1B LoRA with limited boundary counts and stressor range; coefficients are not established across model scales or full-parameter training & Executed fit + scoped law hypothesis \\
\bottomrule
\end{tabularx}
\end{tighttable}
\end{table}

\begin{ScopeBox}
\textbf{Audit rule used throughout.} We avoid the shorthand ``no parameters cross a boundary,'' because boundary-level aggregated deltas are full-dimensional model updates. The precise statement is: no \emph{per-device} parameters, activations, optimizer states, or per-device updates cross a boundary; only boundary-level securely aggregated deltas and bounded metadata do.
\end{ScopeBox}

\section{Problem Setting and Information-Flow Contract}
\label{sec:contract}
We consider $M$ immutable privacy boundaries $\{K_m\}_{m=1}^M$. Device $i \in K_m$ holds local data distribution $\mathcal{D}_i$ and performs local computation, but its device-level model state may not be exported beyond boundary $m$. The global objective decomposes as
\begin{equation}
\min_{\theta\in\R^d} f(\theta) := \sum_{m=1}^M p_m f_m(\theta), \qquad
f_m(\theta) := \sum_{i\in K_m} p_{i|m} f_i(\theta),
\end{equation}
where $p_m \ge 0$, $\sum_m p_m = 1$, $p_{i|m} \ge 0$, and $\sum_{i\in K_m} p_{i|m}=1$.

\echelon{} maintains three levels of state: a global reference $\theta^{(t)}$, boundary references $\theta_m^{(t)}$, and device-local states $\theta_{m,i}^{(t)}$. Devices optimize locally for $H$ inner steps and emit clipped deltas $u_{m,i}^{(t)} \approx \theta_{m,i}^{(t)} - \theta_m^{(t)}$. Boundaries combine buffered pairs $(u_{m,i}^{(t)},w_{m,i}^{(t)})$ into a middle-step update $\Delta_m^{(t)}$. The global plane never observes any per-device message; it only averages boundary deltas to update $\theta^{(t+1)}$.

\begin{table}[H]
\centering
\begin{tighttable}
\caption{Information-flow contract enforced by \echelon{}.}
\label{tab:contract}
\begin{tabularx}{\linewidth}{>{\raggedright\arraybackslash}X >{\raggedright\arraybackslash}X}
\toprule
Allowed to cross a boundary & Forbidden to cross a boundary \\
\midrule
Boundary-level securely aggregated deltas & Per-device parameters \\
Round identifiers, commitments, and $O(1)$ control metadata & Activations or hidden states \\
Aggregate-level timing and byte-count telemetry & Optimizer states (for example, moments) \\
Optional attestation and audit records & Per-device update vectors or membership-exposing payloads \\
\bottomrule
\end{tabularx}
\end{tighttable}
\end{table}

The contract is deliberately stronger operationally than a generic claim that the system is ``privacy preserving.'' It specifies what may be exported, what may not be exported, and therefore what an audit can verify. The operational audit target is correspondingly concrete: cross-boundary payload decoders and byte counters should show no per-device tensors in flight.

\section{Threat Model, Privacy Scope, and Auditability}
\label{sec:privacy-scope}
\subsection{Threat model}
The primary systems adversary is an honest-but-curious boundary coordinator. It follows the protocol, authenticates devices, manages buffering and timeouts, and executes secure aggregation correctly, but it attempts to infer information about individual device contributions from transcripts and metadata. We assume authenticated devices and secure channels. We do not claim protection against active Byzantine behavior, coordinator--device collusion, or traffic-analysis attacks.

\subsection{What is protected per round}
Within a single secure-aggregation invocation, the boundary coordinator learns only an aggregate
\begin{equation}
S_m^{(t)} := \sum_{i\in A_m^{(t)}} u_{m,i}^{(t)},
\end{equation}
where $A_m^{(t)}$ is the authenticated participant set for invocation $t$ and $|A_m^{(t)}|\ge q$ satisfies quorum. Masks and seeds are erased after successful unmasking. Transcripts do not reveal the ordering of contributions, and the global plane does not receive per-device messages. Any two multisets of per-device updates with the same sum induce the same observable view to an honest-but-curious coordinator inside this per-round model.

\begin{figure}[H]
\centering
\includegraphics[width=0.94\linewidth]{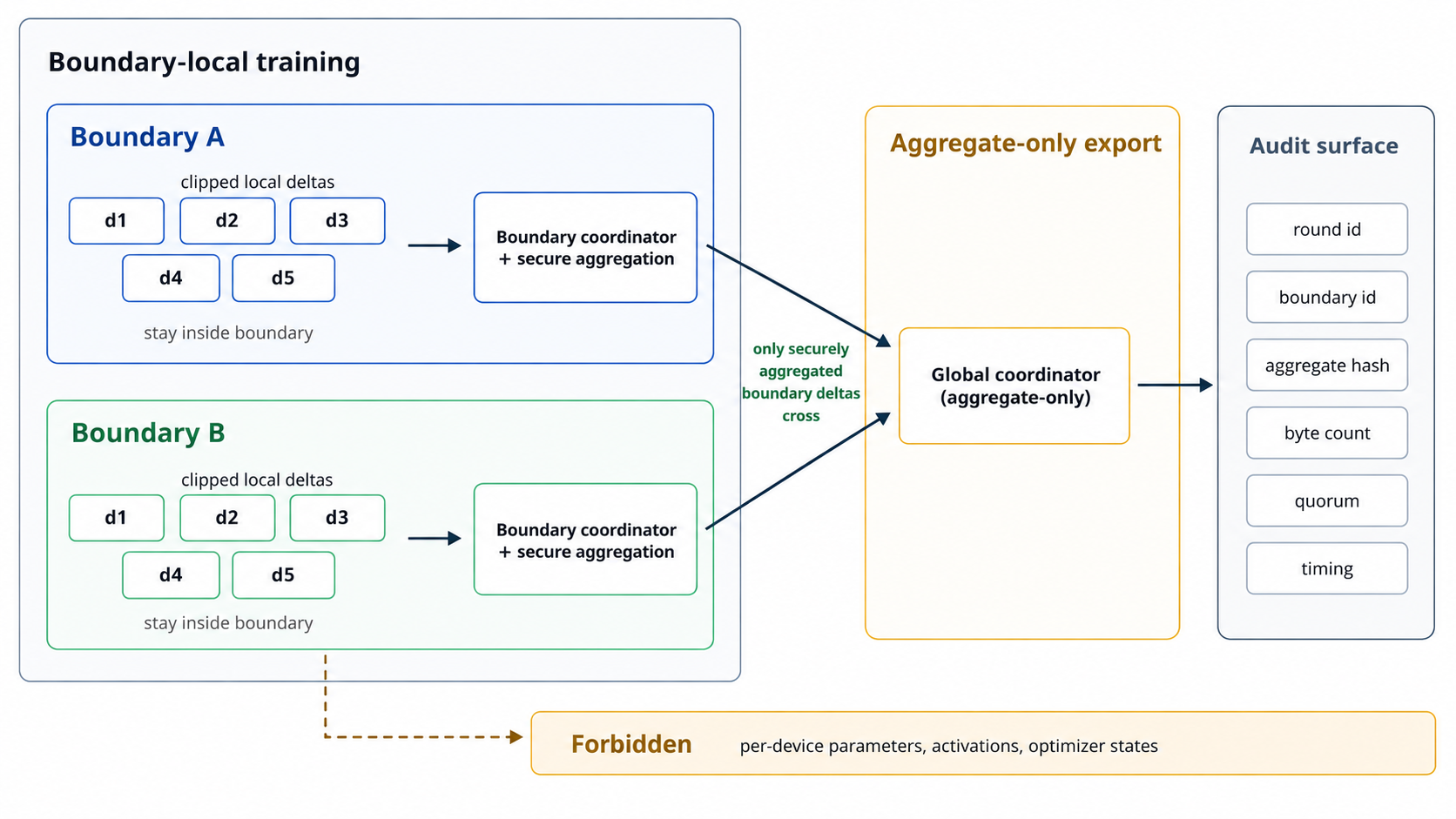}
\caption{Boundary-first architecture and audit surface. Devices optimize and communicate only within their boundary. The global plane sees aggregate-only boundary deltas plus $O(1)$ metadata. The dashed warning path shows flows forbidden by construction.}
\label{fig:architecture}
\end{figure}

\subsection{What is not protected across rounds}
Per-round confidentiality is not longitudinal privacy. Across repeated rounds, an adversarial coordinator could attempt difference-of-sums attacks, reason about small cohorts, or exploit membership churn to narrow uncertainty about an individual's contribution. We therefore treat longitudinal leakage as a systems risk to be bounded operationally rather than hidden behind an overbroad privacy claim.

The current design mitigates, but does not formally eliminate, this risk. It uses random sampling and shuffling within boundaries, bounded churn and fixed-membership windows, minimum quorums for secure aggregation, and avoidance of externally exposed detailed membership lists. These mechanisms reduce the opportunity to engineer informative differences across adjacent rounds, but they are not substitutes for differential privacy or a full leakage proof.

\begin{ScopeBox}
\textbf{What this paper claims.} Device-level model-state non-export by construction; secure-aggregation confidentiality for a single boundary invocation under quorum; auditable aggregate-only exchange.\par
\textbf{What this paper does not claim.} Differential privacy; robustness to active or Byzantine adversaries; immunity to traffic analysis; complete formal protection against longitudinal inference across many rounds.
\end{ScopeBox}

\subsection{Auditability as a systems property}
The architecture exposes a narrow audit surface. Because only aggregate messages may cross a boundary, an auditor can inspect message schemas, byte counts, round identifiers, aggregate hashes, quorum events, and timing logs to confirm that the system obeys the information-flow contract. The value is operational: privacy is encoded into what the system can send, not merely stated as an aspiration.

\section{System Design}
\subsection{Three execution planes}
\begin{figure}[H]
\centering
\includegraphics[width=0.90\linewidth]{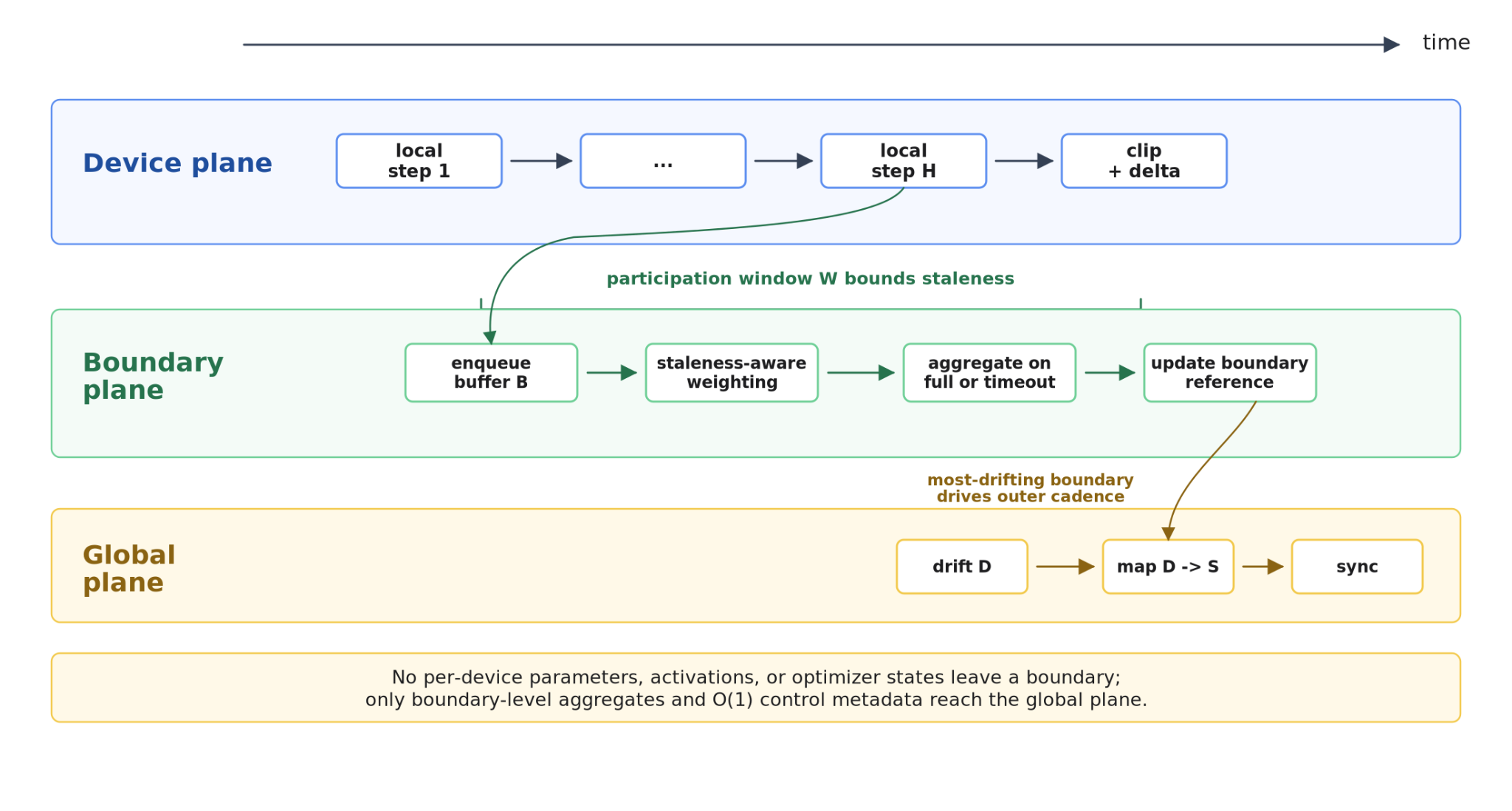}
\caption{Execution flow in \echelon{}. Local device steps produce clipped deltas. Boundaries buffer and securely aggregate semi-asynchronously with staleness-aware weighting and participation windows. The global plane only observes boundary-level drift signals and aggregate deltas; the most-drifting boundary drives outer cadence.}
\label{fig:flow}
\end{figure}

Figures~\ref{fig:architecture} and~\ref{fig:flow} show the three execution planes. The device plane performs local optimization inside a boundary. The boundary plane is the systems layer where buffering, participation windows, clipping, staleness-aware weighting, and secure aggregation are enforced. The global plane is intentionally simple: it receives aggregate-only boundary deltas and mixes them at an outer cadence.

\subsection{Device-local objective}
Each device performs $H$ local steps against a boundary reference. To stabilize non-IID shards, we use a proximal local objective
\begin{equation}
\min_{\theta_{m,i}}\;\mathbb{E}_{x\sim\mathcal{D}_i}\ell(\theta_{m,i};x) + \frac{\mu}{2}\left\|\theta_{m,i}-\bar{\theta}_m\right\|_2^2,
\end{equation}
where $\bar{\theta}_m$ is the current boundary reference or an EMA-smoothed version of it. The proximal term is not introduced as a new algorithmic idea; it is used because aggregate-only exchange amplifies the practical cost of local drift. After $H$ local steps, the device emits a clipped delta $u_{m,i}$ with global $\ell_2$ clipping threshold $C$.

\subsection{Buffered semi-asynchronous aggregation}
Each boundary maintains a buffer of capacity $B$. Aggregation fires when the buffer is full or a timeout $T_{\max}$ expires. Devices that have not participated in the last $W$ middle steps are prioritized, which bounds starvation and limits the age of information. Let $\tau_{m,i}$ denote the age since device $i$ last participated. We use staleness-aware weights
\begin{equation}
w_{m,i}=\exp(-\lambda\tau_{m,i}),
\end{equation}
and form a boundary update from the clipped and weighted contributions. This targets irregular arrivals under WAN delay: the boundary can continue making progress without waiting for perfect synchrony, while stale updates are down-weighted.

\subsection{Drift-aware outer synchronization}
Buffered middle steps reduce boundary-local waiting, but they also allow boundaries to diverge. \echelon{}-DA adds an outer control loop. Each boundary tracks a drift proxy
\begin{equation}
D_{t+1,m} = (1-\beta)\lVert\Delta_{t,m}\rVert_2^2 + \beta D_{t,m},
\end{equation}
and maps it to an integer outer interval
\begin{equation}
S_{t+1,m} = S_{\min} + (S_{\max}-S_{\min})\,\sigma\!\left(\gamma(\vartheta-D_{t+1,m})\right),
\end{equation}
where $\sigma(\cdot)$ is the logistic function. As drift increases, the scheduled interval decreases. The global plane synchronizes when the current step exceeds the smallest boundary interval, $S_t^\star=\min_m S_{t,m}$. In words, the most-drifting boundary drives outer cadence.

\subsection{Implementation notes}
The implementation uses PyTorch for model execution and a boundary microservice for queueing, secure-aggregation orchestration, timeouts, and telemetry. NCCL is used only within a boundary. Compression and quantization are boundary-local implementation choices; the cross-boundary payload remains a masked aggregate plus $O(1)$ metadata. The implementation logs only aggregate or boundary-local counters such as buffer occupancy, secure-aggregation timing, p95/p99 wait, participation gaps, and byte counts.

\begin{algorithm}[H]
\caption{\echelon{}-DA (boundary-scoped buffered semi-asynchronous training).}
\label{alg:echelon}
\footnotesize
\begin{algorithmic}[1]
\Require boundaries $\{K_m\}_{m=1}^M$, buffer capacity $B$, timeout $T_{\max}$, window $W$, local steps $H$, clipping threshold $C$, staleness decay $\lambda$, proximal strength $\mu$, drift EMA $\beta$, interval bounds $[S_{\min},S_{\max}]$
\State Initialize global reference $\theta$, boundary references $\{\theta_m\}$, drift states $\{D_m\}$, and interval states $\{S_m\}$
\For{$t = 1, 2, \ldots, N$}
  \ForAll{boundaries $m$ in parallel}
    \ForAll{devices $i \in K_m$ selected for local work}
      \State Set local state $\theta_{m,i} \leftarrow \theta_m$
      \For{$h = 1$ to $H$}
        \State Take a local gradient step on $\ell(\theta_{m,i};x) + \frac{\mu}{2}\lVert\theta_{m,i} - \bar{\theta}_m\rVert^2$
      \EndFor
      \State Form clipped delta $u_{m,i} \leftarrow \mathrm{clip}(\theta_{m,i}-\theta_m,C)$
      \State Compute staleness weight $w_{m,i} \leftarrow e^{-\lambda \tau_{m,i}}$ and enqueue $(u_{m,i},w_{m,i})$
    \EndFor
    \If{buffer full or timeout expires}
      \State Run secure aggregation on buffered contributions to form $\Delta_m$
      \State Update boundary reference $\theta_m \leftarrow \mathrm{MiddleOpt}(\theta_m,\Delta_m)$
      \State Update drift proxy $D_m \leftarrow (1-\beta)\lVert\Delta_m\rVert_2^2 + \beta D_m$
      \State Update interval $S_m$ via drift controller; enforce participation window $W$
    \EndIf
  \EndFor
  \If{outer step triggered with cadence $S_t^\star = \min_m S_{t,m}$}
    \State Mix aggregate-only boundary deltas to update $\theta$
  \EndIf
\EndFor
\end{algorithmic}
\end{algorithm}

\section{Analytical Support}
The theoretical component of the paper is intended as support for the systems design rather than its primary novelty. Assume each local objective is $L$-smooth, clipped stochastic gradients have bounded second moment, buffering and participation windows induce bounded staleness $\tau_{m,i}\le \tau_{\max}$, and secure aggregation reveals only a quorum-satisfying masked sum. Let the boundary estimator be the clipped, staleness-weighted aggregate actually used by the algorithm.

Here, $L$ denotes the smoothness constant of the local objectives, $\sigma^2$ represents the variance bound of the clipped stochastic gradients, and $\tau_{\max}$ is the maximum delay introduced by semi-asynchronous aggregation. The analysis assumes bounded staleness, clipped updates, and quorum-based secure aggregation under an honest-but-curious coordinator.
\begin{theorem}[Outer descent under bounded delay]
Under the assumptions above, choose a local step size $\eta_m\le 1/(4L)$ and a global step size $\eta\le 1/(2L)$. After $T$ outer steps,
\begin{equation}
\frac{1}{T}\sum_{t=0}^{T-1}\mathbb{E}\left[\lVert\nabla f(\theta^{(t)})\rVert_2^2\right]
\le
\frac{2(f(\theta^{(0)})-f^\star)}{\eta T}+c_1\eta_m(\sigma^2+C^2)+c_2\frac{L^2\tau_{\max}^2}{\lambda},
\end{equation}
for constants $c_1,c_2>0$ independent of $T$.
\end{theorem}

The final term isolates the systems trade-off introduced by semi-asynchrony. Larger maximum delay worsens the bias term; stronger down-weighting and tighter participation control reduce it.

\begin{proposition}[Effect of proximal stabilization and gating]
Under non-IID device distributions, the proximal objective contracts device deviations around the boundary reference and reduces the variance of boundary deltas. Conditioning on alignment with a boundary-level direction estimate increases the expected directional alignment of accepted middle updates at fixed second moment.
\end{proposition}

\begin{remark}
These statements are conservative and are meant to provide intuition for why bounded-delay aggregate-only training can remain stable. The paper's primary contribution is the auditable system design and its empirical characterization; the analysis is included as supporting context rather than as a new theoretical primitive.
\end{remark}

\section{Experimental Methodology}
\begin{ScopeBox}
\textbf{Scope of the current evidence.} The experiments study 1B-parameter LoRA adaptation with sequence length 32, micro-batch size 6, and usually $M=2$ boundaries under controlled WAN emulation using \texttt{tc/netem}. We report the results as evidence for boundary-first aggregate-only adaptation, not as proof of full-scale internet-wide pretraining behavior.
\end{ScopeBox}

\subsection{Datasets, model, and partitioning}
We report results from two corpora selected for complementary goals. English C4 (C4-XS/C4-S) supports budget-matched comparisons and a fixed-sync privacy snapshot; OpenWebText supports workload-realism and WAN-latency stress sweeps where rapid iteration and time-to-target measurements are important. Because absolute perplexity depends strongly on dataset, preprocessing, and training horizon, we compare methods within each regime and label each result block accordingly.

Across all experiments, the model is a 1B-parameter Llama-family variant with LoRA adapters on \texttt{q\_proj} and \texttt{v\_proj}, rank $r=16$, $\alpha=6$, and dropout 0.2 unless otherwise stated. Unless noted, experiments use sequence length 32, micro-batch size 6, gradient accumulation 4, mixed precision, gradient checkpointing, and global clipping $C=1.0$. IID baselines split by tokens; non-IID runs use Dirichlet partitions with $\alpha\in\{0.05,0.1,0.3,1.0\}$ over a proxy host/domain label.

All corpora used in this work are publicly available and are used in accordance with their respective terms. The base model is a Llama-family variant used under its license; we do not redistribute model weights.

\begin{table}[H]
\centering
\begin{tighttable}
\caption{Core experimental regime for the current paper.}
\label{tab:core-regime}
\begin{tabularx}{0.82\linewidth}{>{\raggedright\arraybackslash}p{0.36\linewidth} >{\raggedright\arraybackslash}X}
\toprule
Knob & Default value \\
\midrule
Model & 1B-parameter Llama-family model \\
Adaptation method & LoRA on \texttt{q\_proj}, \texttt{v\_proj}; $r=16$, $\alpha=6$, dropout 0.2 \\
Sequence length & 32 \\
Micro-batch / grad accumulation & 6 / 4 \\
Boundaries & $M=2$ unless noted \\
Boundary controls & $B=4$, $W=4$, $T_{\max}=0.5$ s \\
Staleness / proximal & $\lambda=0.05$, $\mu=0.05$ \\
Drift-aware sync & $S_{\min}=5$, $S_{\max}=100$, $\beta=0.95$ \\
WAN medians & 5, 20, 100 ms with 50\% jitter \\
\bottomrule
\end{tabularx}
\end{tighttable}
\end{table}

\subsection{Baselines, privacy parity, and tuning budget}
We compare against centralized non-private training, FedAvg, FedProx, SCAFFOLD, DiLoCo, and a privacy-parity DiLoCo+SA overlay. All baselines share tokenizer, sequence length, LoRA target modules, learning-rate schedule, clipping, and mixed-precision settings. Privacy-parity overlays are used to compare aggregate-only exchange against methods that otherwise would export model updates across boundaries.

Baseline fairness is particularly sensitive in low-communication training: changing synchronization interval, local-step count, and proximal/staleness coefficients can move a method along a quality--communication frontier. For the Regime BM contest, each method receives a matched hyperparameter-search budget measured in tokens processed by pilot runs on the same C4 slice under the same evaluation cadence. Specifically, before launching the full 24.88M-token evaluations, each method receives an additional 24.88M-token-equivalent pilot budget, executed as $N_{\mathrm{trial}}=8$ pilot trials of 3.11M tokens each, with early stopping allowed. The selected configuration is then re-run for $n_{\mathrm{seeds}}=3$ at the full 24.88M-token budget.

\noindent\textbf{Admissibility under the boundary contract.} Some baselines are included to measure optimization quality under familiar distributed-training assumptions, but they are not all admissible under the target deployment constraint unless augmented with privacy-parity mechanisms. In the intended regime, a method that exports per-device updates or device-level model state across administrative boundaries violates the information-flow contract. We therefore report both standard low-communication baselines and privacy-parity variants, and interpret results along two axes: optimization quality under matched budgets and deployability under the boundary contract.

\section{Results and Empirical Evidence}
The empirical evaluation is organized around three regimes: FS, WR, and BM. These are controlled comparison universes within which absolute perplexity is meaningful and across which it is not.

\begin{table}[H]
\centering
\begin{tighttable}
\caption{Regime reconciliation. Absolute PPL values are only comparable within a regime.}
\label{tab:regimes}
\begin{tabularx}{\linewidth}{>{\raggedright\arraybackslash}p{0.12\linewidth} >{\raggedright\arraybackslash}p{0.20\linewidth} >{\raggedright\arraybackslash}p{0.25\linewidth} >{\raggedright\arraybackslash}p{0.12\linewidth} >{\raggedright\arraybackslash}X}
\toprule
Regime & Corpus & Training horizon & Seeds & Used for \\
\midrule
FS & C4 & 150--250 epochs & 1 & Fixed-sync operating point (single seed; stability check) \\
WR & OpenWebText & 30 epochs & 1 & Workload-realism / WAN stress \\
BM & C4 & 24.88M tokens & 3 & Headline baseline contest \\
\bottomrule
\end{tabularx}
\end{tighttable}
\end{table}

\subsection{Budget-matched baseline contest}
The headline empirical comparison is the seed-averaged Regime BM contest. Table~\ref{tab:bm-main} compares \echelon{} against low-communication DiLoCo-style baselines under a matched-token evaluation budget. All methods use 24.88M tokens; values are reported over $n_{\mathrm{seeds}}=3$ where complete.

\begin{table}[H]
\centering
\begin{tighttable}
\caption{Regime BM (C4): budget-matched baseline contest. Lower validation loss and perplexity are better. Absolute PPL values are only comparable within Regime BM.}
\label{tab:bm-main}
\begin{tabularx}{\linewidth}{>{\raggedright\arraybackslash}Xccccc}
\toprule
Baseline & Final val. loss & PPL & Comm. (GB) & Syncs & Tokens (M) \\
\midrule
\echelon{} & $3.887\pm0.010$ & 48.75 & $18\pm2$ & $12\pm2$ & 24.88 \\
Drift DiLoCo & $4.03\pm0.02$ & 56--57 & $18\pm2$ & $12\pm2$ & 24.88 \\
DiLoCo+SA & $3.95\pm0.01$ & $\sim$52.2 & 35--45 & 15--25 & 24.88 \\
Streaming DiLoCo & $3.95\pm0.01$ & $\sim$52.2 & 30--40 & 30--60 & 24.88 \\
Standard DiLoCo & $3.95\pm0.01$ & $\sim$52.1 & 30--40 & 8--15 & 24.88 \\
\bottomrule
\end{tabularx}
\end{tighttable}
\end{table}

To avoid a single-axis comparison, Table~\ref{tab:bm-budget} reports fixed-budget normalizations by fixing one budget dimension at a time: bytes, wall-clock, tokens, or synchronization count. Under these fixed budgets, \echelon{} is best or tied-best across the four normalizations in this 1B-parameter LoRA adaptation regime.

\begin{table}[H]
\centering
\begin{tighttable}
\caption{Regime BM (C4): fixed-budget validation-loss comparisons derived from the budget-matched runs. Assumed budgets are fixed-bytes = 18 GB, fixed-wall-clock = 27,000 s, fixed-token = 24.88M, and fixed-sync-count = 12. Lower is better.}
\label{tab:bm-budget}
\begin{tabularx}{\linewidth}{>{\raggedright\arraybackslash}Xcccc}
\toprule
Baseline & Fixed-bytes & Fixed-wall-clock & Fixed-token & Fixed-sync-count \\
\midrule
\echelon{} & 3.89 & 3.89 & 3.89 & 3.89 \\
Drift DiLoCo & 4.03 & 4.02 & 4.03 & 4.03 \\
DiLoCo+SA & 4.02--4.05 & 3.96--4.00 & 3.95 & 3.98--4.02 \\
Streaming DiLoCo & 4.03--4.06 & 3.96--4.00 & 3.95 & 4.01--4.05 \\
Standard DiLoCo & 4.00--4.04 & 3.95--3.98 & 3.95 & 3.97--4.01 \\
\bottomrule
\end{tabularx}
\end{tighttable}
\end{table}

\subsection{Workload realism and WAN stress}
Regime WR uses OpenWebText with Llama-3.2-1B on 6 L40S-class GPUs, DiLoCo sync interval $H=2$, micro-batch size 6, and sequence length 32 unless noted. Table~\ref{tab:wr-footprint} shows a trainable-footprint sweep. More trainable parameters monotonically reduce loss but lower throughput; the boundary pipeline is agnostic to the LoRA topology.

\begin{table}[H]
\centering
\begin{tighttable}
\caption{Regime WR (OpenWebText): trainable footprint sweep (2 boundaries, no WAN, IID, 30 epochs).}
\label{tab:wr-footprint}
\begin{tabularx}{\linewidth}{>{\raggedright\arraybackslash}Xrrrr}
\toprule
LoRA Config & Tune Params & Val Loss & PPL & Tok/s \\
\midrule
\texttt{q/v} r16 QLoRA & 3.4 M & 3.2305 & 25.19 & $\sim$2160 \\
\texttt{q/v} r64 QLoRA & 13.6 M & 3.1172 & 22.58 & 2085 \\
all-attn r16 QLoRA & 6.8 M & 2.8809 & 17.83 & 1817 \\
attn+MLP r16 QLoRA & 12.6 M & 2.2109 & 9.12 & 1506 \\
\bottomrule
\end{tabularx}
\end{tighttable}
\end{table}

\begin{table}[H]
\centering
\begin{tighttable}
\caption{Regime WR (OpenWebText): stress resilience under emulated WAN and non-IID data (\texttt{q/v} LoRA r16, QLoRA, 2 boundaries, 30 epochs).}
\label{tab:wr-stress}
\begin{tabular}{lllrrr}
\toprule
Latency & WAN & Non-IID & Val Loss & PPL & Tok/s \\
\midrule
0 ms & none & IID & 3.2305 & 25.19 & $\sim$2160 \\
0 ms & none & severe & 3.2988 & 27.08 & 2160 \\
50 ms & bursty & severe & 3.2988 & 27.08 & 2176 \\
200 ms & none & IID & 3.3027 & 27.19 & 2139 \\
\bottomrule
\end{tabular}
\end{tighttable}
\end{table}

\begin{figure}[H]
\centering
\includegraphics[width=0.78\linewidth]{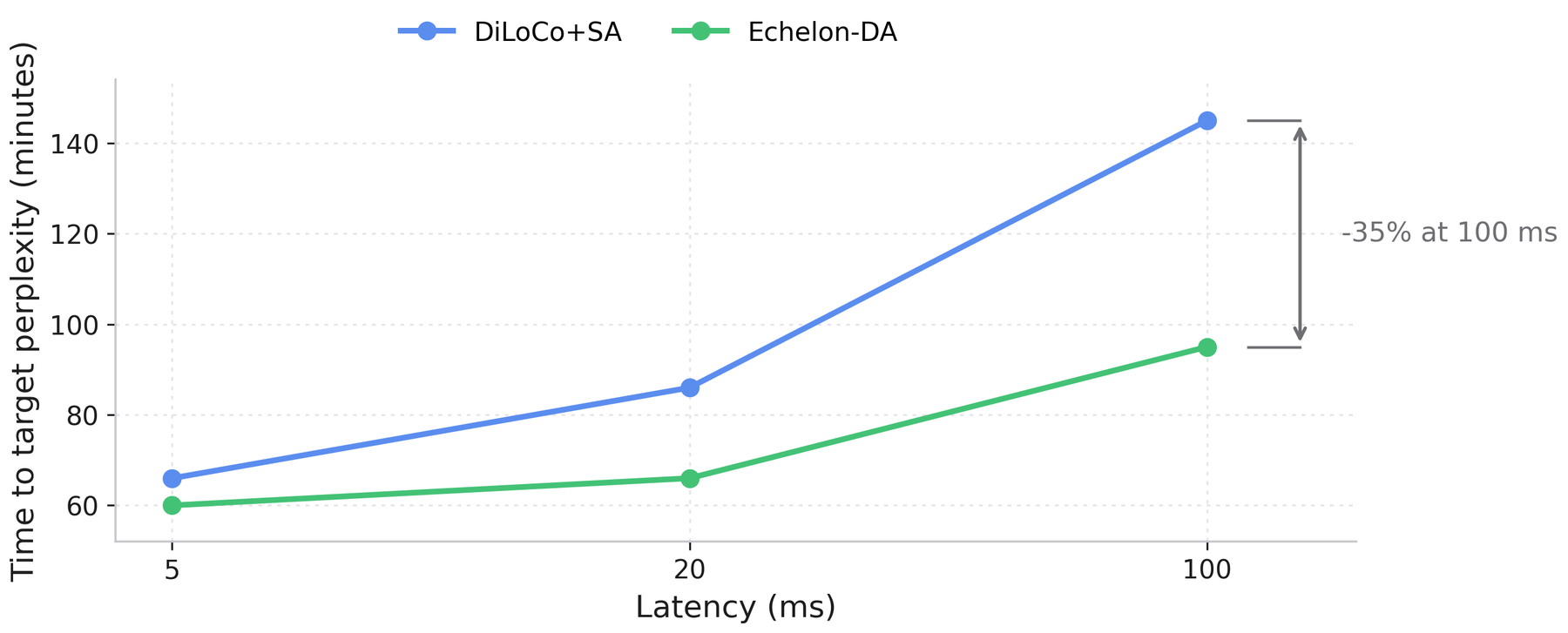}
\caption{Regime WR (OpenWebText): latency sensitivity under emulated WAN delay. We plot time-to-target perplexity versus median WAN latency for \echelon{}-DA and DiLoCo+SA; the gap widens at 100 ms. See Table~\ref{tab:wr-latency-time}.}
\label{fig:latency}
\end{figure}

Validation loss degrades by at most 2.2\% under 200 ms emulated WAN latency or severe non-IID partitioning in this block, and throughput remains within 1\% of the unstressed baseline. Figure~\ref{fig:latency} shows the corresponding time-to-target sensitivity under latency.

\begin{table}[H]
\centering
\begin{tighttable}
\caption{Regime WR (OpenWebText): time-to-target under WAN latency. Values (minutes) correspond to the points in Figure~\ref{fig:latency}.}
\label{tab:wr-latency-time}
\begin{tabular}{lccc}
\toprule
Method & 5\,ms & 20\,ms & 100\,ms \\
\midrule
DiLoCo+SA & 66 & 86 & 145 \\
\echelon{}-DA & 60 & 66 & 95 \\
\bottomrule
\end{tabular}
\end{tighttable}
\end{table}

\begin{table}[H]
\centering
\begin{tighttable}
\caption{Regime WR (OpenWebText): quality--wall-clock scorecard underlying Figure~\ref{fig:scorecard}. Methods are sorted by time-to-target (minutes); lower is better.}
\label{tab:wr-scorecard}
\begin{tabular}{lcc}
\toprule
Method & Time to target (min) & Validation PPL \\
\midrule
\echelon{}-DA & 52 & 25.10 \\
\echelon{}-DA+SA & 53 & 25.10 \\
Central & 60 & 25.00 \\
\echelon{} & 65 & 25.10 \\
SCAFFOLD & 72 & 25.20 \\
FedProx & 75 & 25.30 \\
DiLoCo & 78 & 25.40 \\
FedAvg & 82 & 25.50 \\
DiLoCo+SA & 85.5 & 25.50 \\
\bottomrule
\end{tabular}
\end{tighttable}
\end{table}

\begin{figure}[H]
\centering
\includegraphics[width=0.86\linewidth]{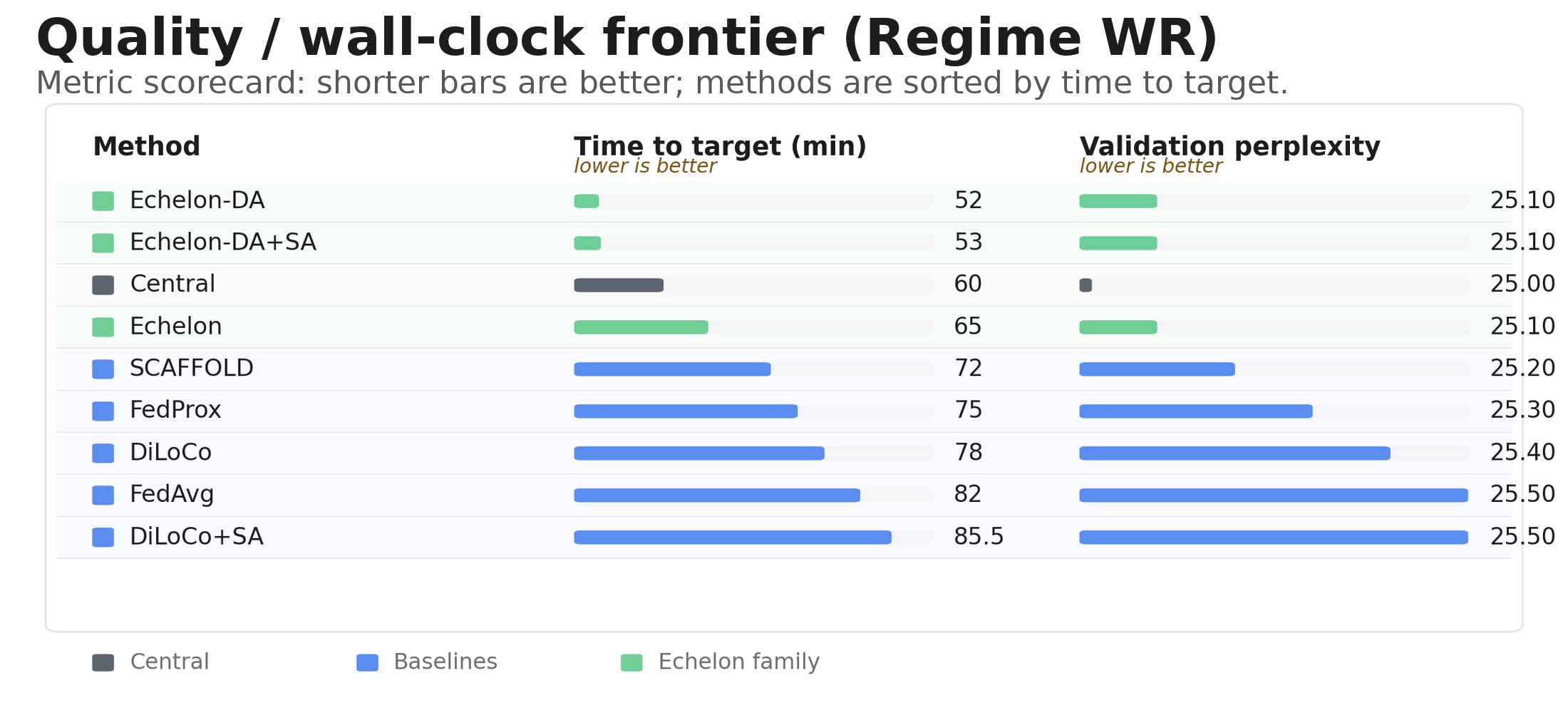}
\caption{Regime WR (OpenWebText): metric scorecard summarizing time-to-target and validation perplexity across baselines under the workload-realism setup. Methods are sorted by time-to-target; shorter bars are better. Underlying values are listed in Table~\ref{tab:wr-scorecard}.}
\label{fig:scorecard}
\end{figure}

\subsection{Drift-controller diagnostics}
We instrument the drift-aware controller and log, at each controller tick $t$: (i) the per-boundary drift proxy $D_{t,m}$, (ii) the instantaneous boundary-update magnitude $\lVert\Delta_{t,m}\rVert_2^2$, (iii) the scheduled outer interval $S_{t,m}$, and (iv) the resulting global cadence $S_t^\star=\min_m S_{t,m}$.

\begin{table}[H]
\centering
\begin{tighttable}
\caption{Representative drift-controller trace. $D_{t,m}$ is the EWMA drift proxy; $S_{t,m}$ is the scheduled outer interval; $S_t^\star$ is the global cadence.}
\label{tab:controller}
\begin{tabular}{rrrrrrrrr}
\toprule
$t$ & $D_{t,0}$ & $D_{t,1}$ & $\|\Delta_{t,0}\|_2^2$ & $\|\Delta_{t,1}\|_2^2$ & $S_{t,0}$ & $S_{t,1}$ & $S_t^\star$ & driver \\
\midrule
0 & 0.13 & 0.22 & 2.61 & 4.41 & 99 & 98 & 98 & 1 \\
1 & 0.25 & 0.43 & 2.45 & 4.71 & 99 & 97 & 97 & 1 \\
2 & 0.36 & 0.64 & 2.49 & 4.62 & 99 & 94 & 94 & 1 \\
3 & 0.47 & 0.83 & 2.51 & 4.42 & 99 & 89 & 89 & 1 \\
8 & 0.92 & 1.59 & 2.38 & 4.50 & 98 & 65 & 65 & 1 \\
15 & 1.74 & 2.83 & 2.88 & 5.12 & 96 & 21 & 21 & 1 \\
16 & 1.80 & 2.94 & 2.91 & 4.71 & 95 & 17 & 17 & 1 \\
17 & 1.86 & 3.03 & 2.98 & 4.85 & 94 & 14 & 14 & 1 \\
18 & 1.82 & 2.98 & 2.27 & 4.21 & 94 & 16 & 16 & 1 \\
19 & 1.88 & 3.12 & 2.95 & 5.65 & 93 & 12 & 12 & 1 \\
\bottomrule
\end{tabular}
\end{tighttable}
\end{table}

The trace is consistent with the mechanism: boundary 1 is more divergent, maps to a tighter schedule, and determines global cadence. The policy is responsive rather than monotone: when boundary 1's instantaneous update magnitude and drift proxy dip at $t=18$, the scheduled interval relaxes from 14 to 16 before tightening again.

\subsection{Fixed-sync operating-point snapshot (single seed)}
\label{sec:regime-fs}
Table~\ref{tab:fs} reports a long-horizon fixed-sync operating-point snapshot in Regime FS (C4), run with a single seed to sanity-check stability under the aggregate-only contract. \echelon{} + Privacy (LoRA) reaches validation perplexity 9.63 (loss 2.26), while Drift DiLoCo reaches 12.20 (loss 2.50). We include this block as a feasibility/stability check; the headline multi-seed comparisons are in Regime BM.

\begin{table}[H]
\centering
\begin{tighttable}
\caption{Regime FS (C4), single seed. Long-horizon fixed-sync operating-point snapshot (stability check).}
\label{tab:fs}
\begin{tabularx}{\linewidth}{>{\raggedright\arraybackslash}Xrrrrr}
\toprule
Method & Val. PPL & Val. Loss & Train PPL & Epochs & Tok/s \\
\midrule
\echelon{} + Privacy (LoRA) & 9.63 & 2.26 & 5.40 & 150 & 988 \\
Drift DiLoCo & 12.20 & 2.50 & 6.78 & 200 & 528 \\
\bottomrule
\end{tabularx}
\end{tighttable}
\end{table}

A privacy ablation in the same regime suggests that enforcing the contract does not induce a large quality penalty in this setting: the privacy-constrained \echelon{} run (9.63 PPL) is close to a non-private \echelon{} LoRA run (9.86 PPL). We treat this as a diagnostic rather than a general claim, since the block is single-seed and tuned for an operating point.

\subsection{Candidate effective-divergence law}
We evaluate whether controller telemetry admits a one-dimensional empirical law for the quality--communication frontier. The evaluation suite spans boundary counts $M\in\{2,3\}$, round-trip latencies $t_{\mathrm{RTT}}\in\{0,100\,\mathrm{ms}\}$, and three non-IID severities (Dirichlet $\alpha\in\{1.0,0.1,0.01\}$ inter-group), alongside a centralized $M=1$ baseline ($\Lval=3.27$). For each trajectory, $\Beff$ is computed from per-boundary divergence telemetry and adaptive synchronization intervals, using EMA decay $\beta=0.95$.

\begin{figure}[H]
\centering
\includegraphics[width=0.86\linewidth]{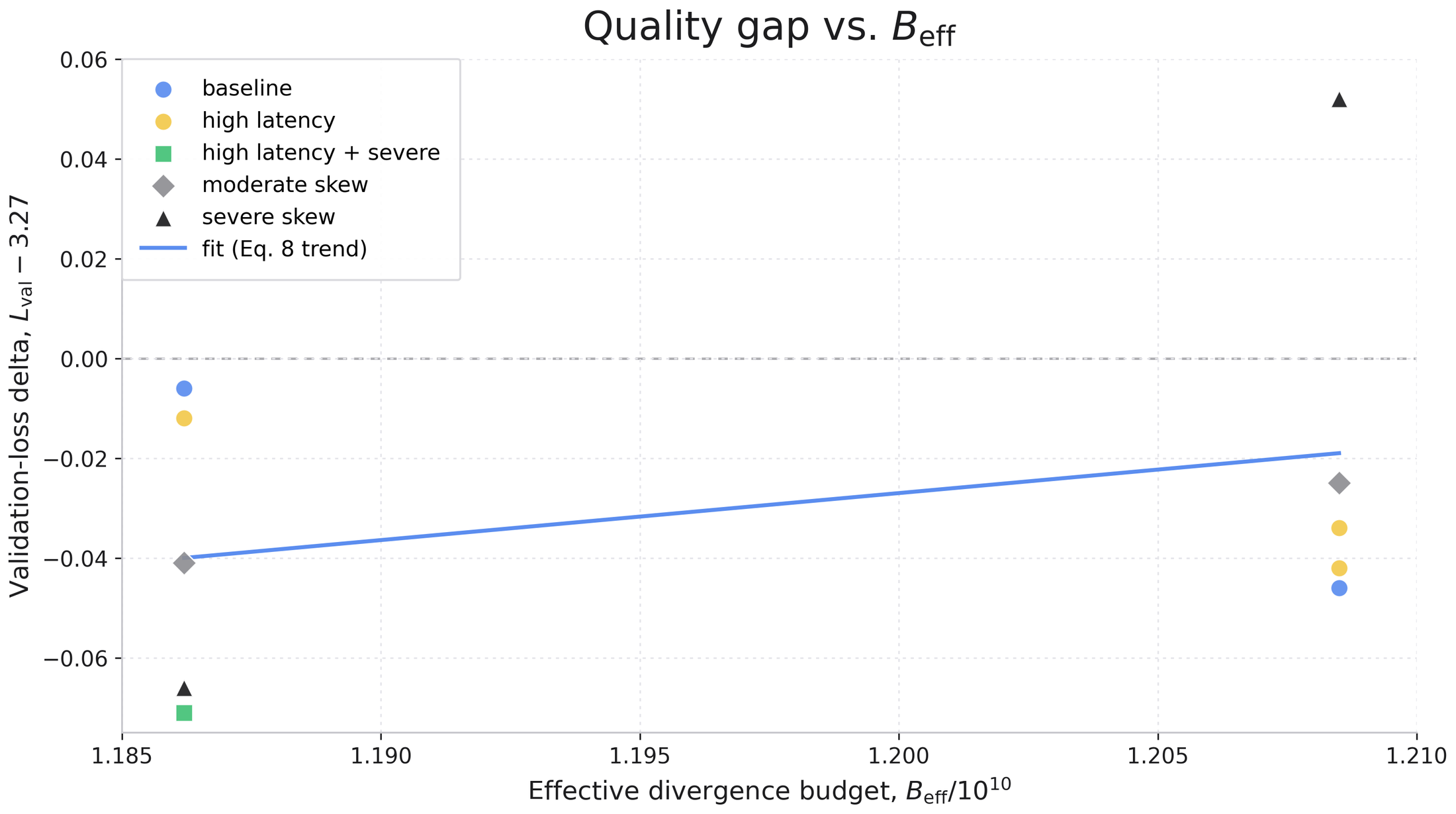}
\caption{Quality gap versus normalized $\Beff$. Validation-loss delta (relative to the $M=1$ baseline of 3.27; negative is better) is plotted against $\widetilde{B}_{\mathrm{eff}}=\Beff/10^{10}$ for the stressor sweep spanning $M\in\{2,3\}$, variable latency, and non-IID severity.}
\label{fig:beff-quality}
\end{figure}

\begin{figure}[H]
\centering
\includegraphics[width=0.86\linewidth]{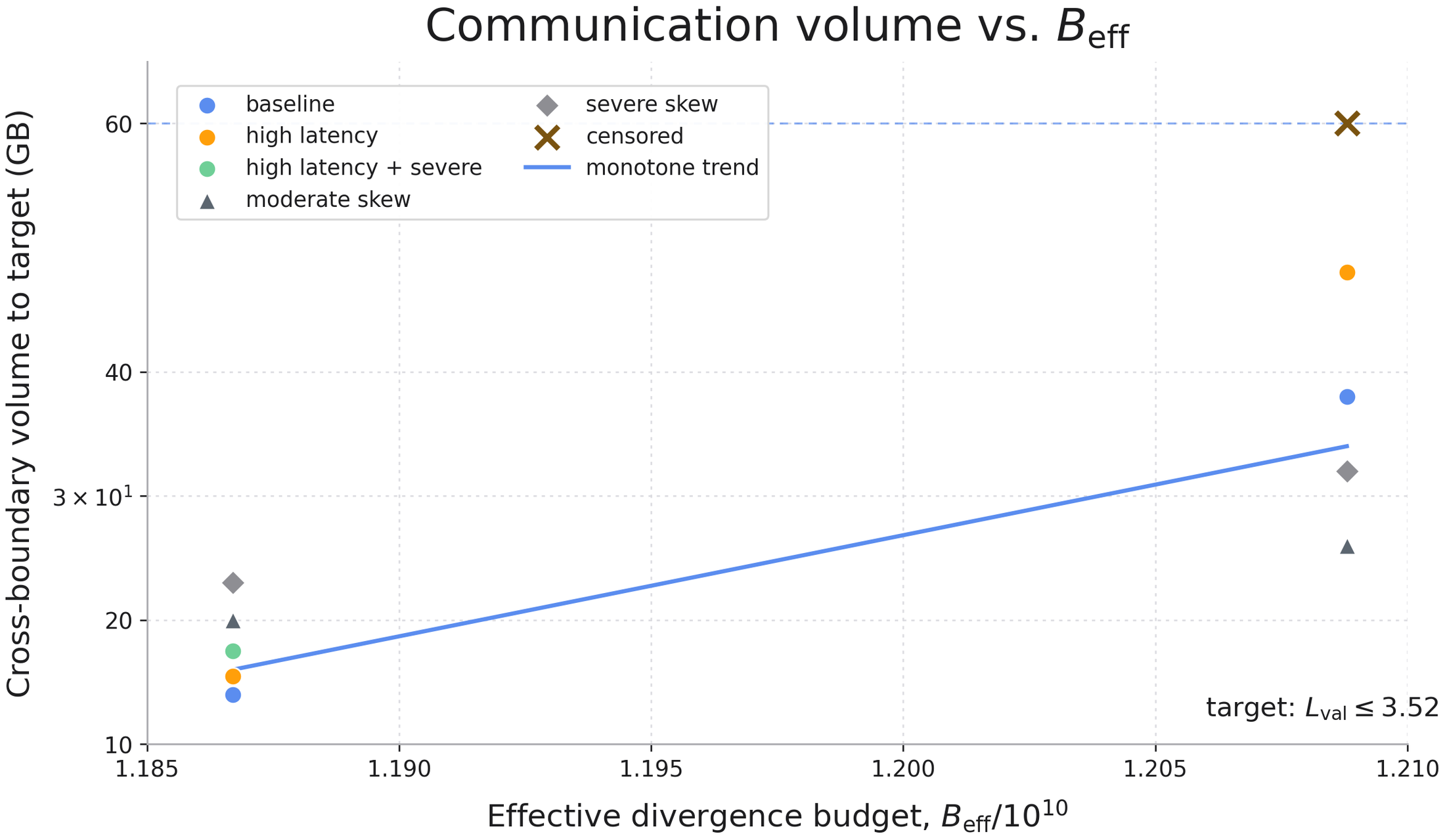}
\caption{Communication volume versus normalized $\Beff$. Cross-boundary bytes required to reach $\Lval\le3.52$ (1.10x baseline) are plotted against $\widetilde{B}_{\mathrm{eff}}=\Beff/10^{10}$. Censored runs are indicated with muted amber crosses.}
\label{fig:beff-comm}
\end{figure}

Let $B_0=10^{10}$ be a fixed normalization constant matching the scale used in Figures~\ref{fig:beff-quality} and~\ref{fig:beff-comm}, and define the normalized divergence coordinate
\begin{equation}
\widetilde{B}_{\mathrm{eff}} = \Beff / B_0.
\end{equation}
Over the evaluated operating window, we fit the empirical relation
\begin{equation}
\Delta L_{\mathrm{val}} = a\cdot \widetilde{B}_{\mathrm{eff}}^{\,b} + c + \epsilon,
\label{eq:beff-law}
\end{equation}
where $\Delta L_{\mathrm{val}}$ is the validation-loss gap relative to the $M=1$ baseline and $\epsilon$ denotes residual variation not captured by the one-scalar law.

We call Equation~\ref{eq:beff-law} a candidate effective-divergence law because it links a boundary-induced systems quantity, $\Beff$, to the observed quality gap under a fixed adaptation family. The claim is deliberately scoped. We do not assert that the coefficients are universal across model scales, context lengths, architectures, or full-parameter training. Rather, the falsifiable claim is that within a specified boundary-local training stack and operating window, $\Beff$ provides a compact predictor of the quality--communication frontier. Figures~\ref{fig:beff-quality} and~\ref{fig:beff-comm} show the quality and communication views.

\begin{table}[H]
\centering
\begin{tighttable}
\caption{Predictive validation for the candidate effective-divergence law. ``Pr($\Beff$ better)'' denotes the bootstrap probability that the one-scalar $\Beff$ model achieves lower RMSE than the baseline.}
\label{tab:beff-validation}
\begin{tabularx}{\linewidth}{>{\raggedright\arraybackslash}Xcccc}
\toprule
Model & CV RMSE & Bootstrap mean & 95\% CI & Pr($\Beff$ better) \\
\midrule
Normalized $\Beff$ law & 0.039 & 0.039 & [0.033, 0.044] & --- \\
Linear-Latency & 0.055 & 0.054 & [0.037, 0.069] & 98.5\% \\
Linear-Boundary Count & 0.056 & 0.055 & [0.040, 0.071] & 99.4\% \\
Linear-$\alpha$ (non-IID) & 0.056 & 0.055 & [0.040, 0.070] & 99.5\% \\
Linear-Bytes & 0.040 & 0.039 & [0.032, 0.046] & 75.9\% \\
Multivariate Linear (all 4) & 0.038 & 0.038 & [0.032, 0.043] & 38.7\% \\
\bottomrule
\end{tabularx}
\end{tighttable}
\end{table}

The normalized $\Beff$ law should be read as a parsimonious predictor rather than as a universal scaling law. In Table~\ref{tab:beff-validation}, the one-scalar $\Beff$ fit improves over raw latency, boundary count, and non-IID severity baselines, and is competitive with linear-bytes and multivariate linear models. Its advantage is interpretability: $\Beff$ is computed from the same boundary-level telemetry that drives the controller, so it summarizes the divergence actually experienced by the system rather than treating latency, boundary count, bytes, and skew as independent external covariates.

The current evidence is not sufficient to claim universality. A stronger law would require validation across model scales, context lengths, task families, boundary counts, and full-parameter training regimes. We therefore present $\Beff$ as a candidate empirical law for boundary-aware adaptation in the evaluated regime, with clear falsification criteria for future work.

\subsection{Privacy red-team and real-WAN audit}
We evaluate the system against an honest-but-curious global coordinator observing all boundary aggregates $\Delta_m^{(t)}$ and participation metadata. The red-team probes three adversarial vectors: difference-of-sums, membership-churn, and small-cohort reconstruction.

\begin{figure}[H]
\centering
\includegraphics[width=\linewidth]{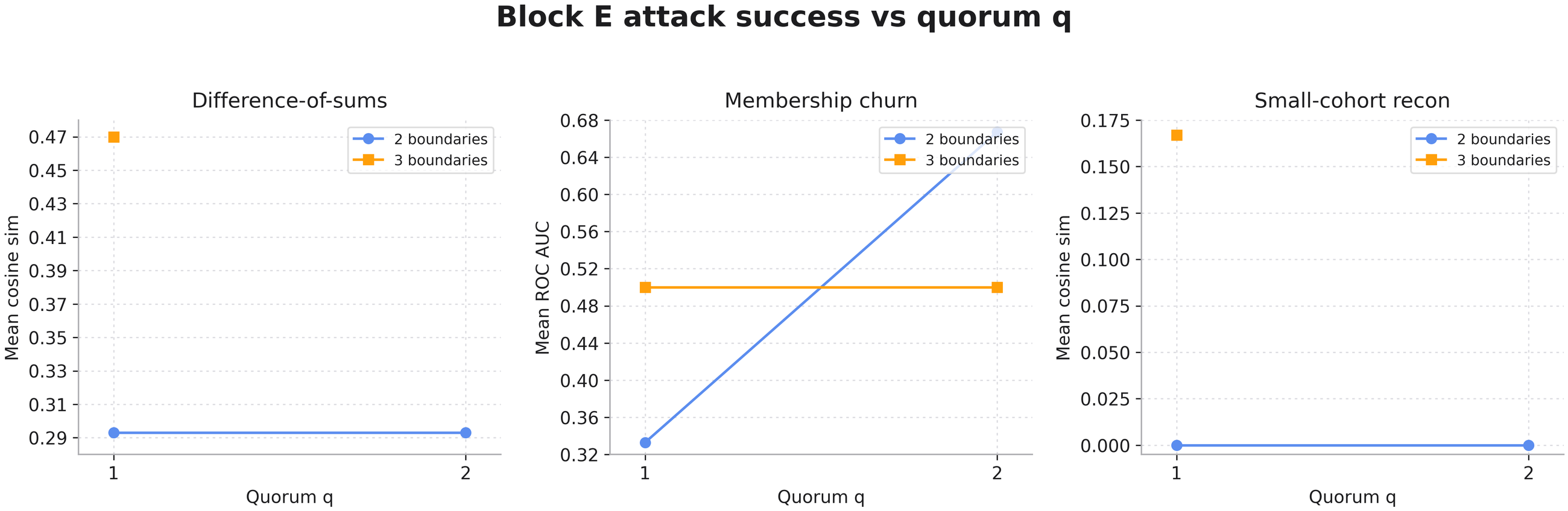}
\caption{Privacy red-team leakage markers under the tested buffer and quorum settings. When the buffer covers the whole boundary, the tested difference-of-sums and small-cohort reconstruction attacks become algebraically undefined because the required distinct aggregate equations are absent.}
\label{fig:redteam}
\end{figure}

\begin{figure}[H]
\centering
\includegraphics[width=\linewidth]{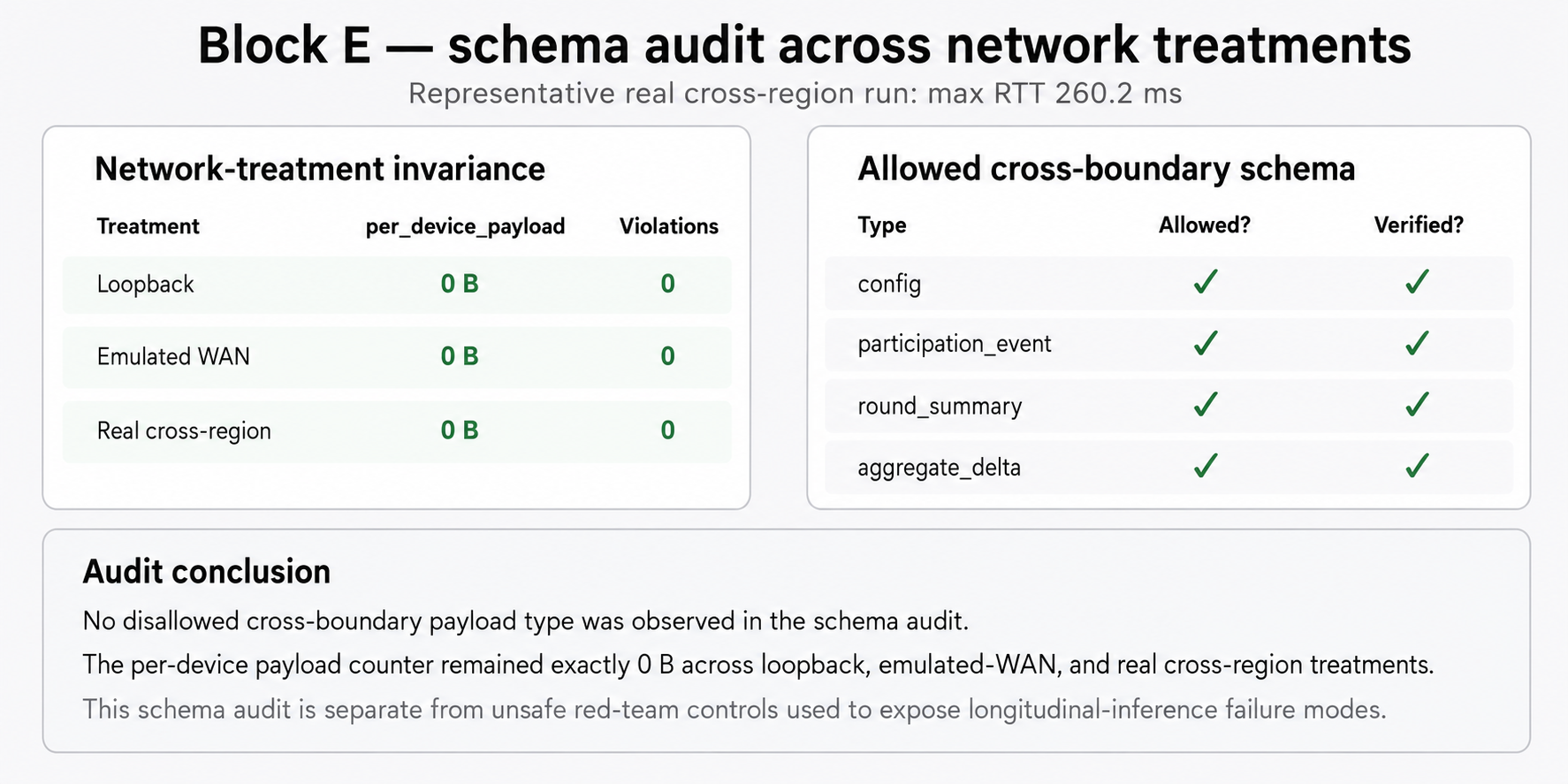}
\caption{Operational audit trace under the tested cross-region conditions. The schema check confirms that \texttt{per\_device\_payload} remains exactly zero bytes, while aggregate and control metadata remain visible to the audit surface.}
\label{fig:realwan-audit}
\end{figure}

\begin{table}[H]
\centering
\begin{tighttable}
\caption{Privacy red-team: worst-case performance. $k:=\max(|\mathcal{B}|-B,0)$ denotes buffer deficit. ``---'' denotes cases where the stated algebraic attack is mathematically undefined under the listed quorum and buffering assumptions.}
\label{tab:redteam}
\begin{tabular}{lrrrrrrrrrr}
\toprule
& & & & & \multicolumn{2}{c}{DoS} & \multicolumn{2}{c}{MC} & \multicolumn{2}{c}{SC} \\
\cmidrule(lr){6-7}\cmidrule(lr){8-9}\cmidrule(lr){10-11}
Run ID & $q$ & $B$ & $|\mathcal{B}|$ & $k$ & Mean & Max & Mean & Max & Mean & Max \\
\midrule
2b\_q1\_b2 & 1 & 2 & 2 & 0 & 0.293 & 0.730 & 0.333 & 0.500 & 0.000 & 0.000 \\
2b\_q2\_b2 & 2 & 2 & 2 & 0 & 0.293 & 0.730 & 0.667 & 1.000 & 0.000 & 0.000 \\
3b\_q1\_b1 & 1 & 1 & 3 & 2 & 0.470 & 1.000 & 0.500 & 1.000 & 0.167 & 1.000 \\
3b\_q2\_b2 & 2 & 2 & 3 & 1 & --- & --- & 0.500 & 0.500 & --- & --- \\
pilot\_q2\_b4 & 2 & 4 & 2 & 0 & --- & --- & 0.500 & 0.500 & --- & --- \\
\bottomrule
\end{tabular}
\end{tighttable}
\end{table}

When the buffer covers every device in the boundary, the structural preconditions for difference-of-sums and small-cohort reconstruction are eliminated by construction. Membership-churn remains a separate vector and must be addressed with metadata minimization and fixed-membership windows.

To verify that the coordinator never observes per-device payloads, we perform a schema-level audit of the operational trace. Across all evaluated runs, the \texttt{per\_device\_payload} count was exactly zero bytes, and zero schema violations were flagged. We validate these properties under loopback, emulated intercontinental WAN, and real-world cross-region operation. The real-world treatment used servers across seven cloud regions and exercised the coordinator protocol end-to-end between region endpoints, with measured RTTs up to 260.2 ms.

\begin{table}[H]
\centering
\begin{tighttable}
\caption{Environmental invariance of the audit surface. Real-WAN figures include cross-region server endpoints across 7 global cloud regions (max RTT 260.2 ms).}
\label{tab:env-audit}
\begin{tabularx}{0.78\linewidth}{>{\raggedright\arraybackslash}Xcc}
\toprule
Treatment & Per-device payload observed & Audit violations \\
\midrule
No WAN (Loopback) & 0 B & 0 \\
Emulated WAN (100 ms) & 0 B & 0 \\
Real Cross-Region & 0 B & 0 \\
\bottomrule
\end{tabularx}
\end{tighttable}
\end{table}

\paragraph{Boundary of the current claim.} The present results should be read as evidence that aggregate-only, auditable adaptation is feasible and competitive in the evaluated 1B-parameter LoRA regimes. They should not be read as evidence for full-parameter internet-scale pretraining, universal privacy protection, or a model-scale-independent scaling law. The broader hypothesis suggested by the results is that, once device-level model-state export is disallowed, the relevant systems frontier is governed by a joint tradeoff among divergence, synchronization, communication, and auditability. $\Beff$ is our first attempt to summarize that frontier in a single measurable quantity.

\section{Limitations}
The paper has clear boundaries.
\begin{itemize}
\item \textbf{Scale.} We study 1B-parameter LoRA adaptation at sequence length 32, typically with $M=2$ boundaries. Long-context, larger-model, and full-parameter scaling remain open.
\item \textbf{WAN realism.} Most quality-oriented WAN results are based on controlled \texttt{tc/netem} emulation. The operational audit includes real cross-region protocol execution, but real-WAN quality validation remains an important next step.
\item \textbf{Privacy scope.} We enforce non-export of device-level state and protect per-round aggregates under quorum, but we do not claim differential privacy, malicious-adversary robustness, or complete protection against longitudinal inference.
\item \textbf{Evidence regime.} Regime FS and WR are single-seed snapshots. The headline empirical comparison is the Regime BM three-seed budget-matched contest.
\item \textbf{Effective-divergence law status.} $\Beff$ is presented as a candidate empirical law for the current training stack and operating window. The manuscript does not establish a universal scaling law. Validating or refuting the law across larger models, longer contexts, more boundaries, full-parameter training, and additional task families remains open.
\end{itemize}

\section{Reproducibility and Artifacts}
This source package is designed for Overleaf and for independent reproduction of the paper's figures and tables. It contains the manuscript source, style file, figure assets, and Decompute mark used for typesetting. Claims involving implementation traces, pilot tuning logs, privacy red-team scripts, and network audit logs should be considered supported by the manuscript record unless those artifacts are uploaded as separate files in the public record.

The experiments use public corpora and a Llama-family base model under its license; no model weights, raw text, raw per-node gradients, or per-device payloads are redistributed in this package. The audit-relevant measurements in the manuscript are aggregate-level byte counts, schema outcomes, and boundary/global telemetry.

\section{Conclusion}
\echelon{} reframes cross-boundary language-model adaptation as an information-flow problem. By enforcing non-export of device-level model state and restricting cross-boundary communication to auditable aggregates, the system enables practical adaptation in regimes where ordinary cross-site model exchange may be inadmissible. The current evidence supports a scoped claim: in a 1B-parameter LoRA regime, aggregate-only coordination can remain competitive under matched budgets, preserve practical throughput under WAN and non-IID stress, and expose an audit surface on which zero per-device payload bytes were observed across tested network treatments.

The results also suggest a broader systems hypothesis: boundary-aware adaptation is governed by an effective divergence budget that couples drift, synchronization, communication, and heterogeneity. We present $\Beff$ as a candidate empirical law for the evaluated operating window, not as a universal scaling law. The remaining work is to scale the validation, release a reusable benchmark and audit suite, pursue external verification of the audit surface, and test whether the effective-divergence law remains predictive across larger models, longer contexts, more boundaries, and full-parameter regimes.

\clearpage
\appendix
\section{Related Work}
\noindent\textbf{Distributed and cross-region LLM training.} Megatron-LM, ZeRO, and DeepSpeed optimized training on tightly coupled clusters, where model-state exchange is fundamental to the design\cite{shoeybi2019megatron,rajbhandari2020zero,rasley2020deepspeed}. More recent low-communication LLM systems such as DiLoCo, OpenDiLoCo, Streaming DiLoCo, and the DiLoCo scaling-law study show that useful training can survive infrequent synchronization, real cross-region deployment, overlapping communication, and larger model scales\cite{douillard2023diloco,jaghouar2024opendiloco,douillard2025streaming,charles2025dilocoscaling}. \echelon{} targets a different axis of the design space: not only low communication, but a hard boundary rule that forbids exporting device-level model state across administrative domains.

\noindent\textbf{Federated optimization and federated LLMs.} FedAvg, FedProx, and SCAFFOLD established the modern vocabulary for communication-efficient learning under heterogeneity\cite{mcmahan2017fedavg,li2020fedprox,karimireddy2020scaffold}. Recent frameworks such as OpenFedLLM broadened the evaluation surface for decentralized LLM training and post-training\cite{ye2024openfedllm}. \echelon{} borrows from this literature, especially proximal stabilization and control for heterogeneity, but its central contribution is architectural: the privacy boundary is explicit and the cross-boundary payload is aggregate-only and auditable.

\noindent\textbf{Secure aggregation, confidentiality, and attack surfaces.} Bonawitz-style secure aggregation provides the cryptographic primitive for per-round aggregate confidentiality\cite{bonawitz2017secagg}. Recent systems work has pushed toward stronger operational guarantees through confidential computation and verifiable secure aggregation, while attack papers have shown that secure aggregation alone does not eliminate risks such as model inconsistency or leakage from aggregated gradients\cite{eichner2024cfc,pasquini2022eluding,wang2024breaking,pu2025janus}. \echelon{} is deliberately scoped within this reality: it claims auditable aggregate-only exchange and per-round confidentiality under an honest-but-curious coordinator, not full protection against malicious or longitudinal attacks.

\noindent\textbf{Parameter-efficient adaptation.} LoRA and related methods make boundary-local adaptation much easier to study in realistic hardware budgets\cite{hu2022lora}. The present paper therefore focuses on language-model adaptation rather than claiming evidence for fully general pretraining.

\end{document}